# Molecular Plasmon Hybridizition in Olefin Chains

Nan Gao[1], Guodong Zhu[1], Yingzhou Huang[2] and Yurui Fang[1, *]

[1.]*Key Laboratory of Materials Modification by Laser, Electron, and Ion Beams (Ministry of Education); School of Physics, Dalian University of Technology, Dalian 116024, P.R. China.*

[2]*State Key Laboratory of Coal Mine Disaster Dynamics and Control, Chongqing University, Chongqing 400044, China*

*Corresponding authors:* yrfang@dlut.edu.cn *(Y.F.)*

**Abstract**

With the continuous emergence of molecular and cluster devices or systems, the relationship between the plasmonic properties of multiple clusters and molecular interactions and the properties of the original single cluster or molecule becomes more and more important. Similar to plasmonic nanoparticle hybridization, there is also a hybrid phenomenon between two molecules with plasmon excitation modes. Using linear response time−dependent density functional theory (LR−TDDFT) and real−time propagation time−dependent density functional theory (RT−TDDFT) and combining the plasmonicity index (PI) and the transition contribution maps (TCM) methods we identify the plasmon excitation mode in the small molecular olefin chains with −OH and −$NH_2$ groups and analyze the hybridization characteristics using charge transitions. The results show that for the plasmons in molecules, there are also plasmon hybridization mechanism exist when the two molecules coupling together. The TCM analysis shows that the plasmon modes and hybridization is a result of coexist of collective and single particle excitation. When there is extra charge depose in the molecules, as the electrons can moving in the whole molecules, the plasmon mode becomes stronger and the individual properties of the molecules maintains in the coupling. The study paves a way for molecule plasmon and the physics picture when the molecules are coupled together.

# 1. Introduction

Plasmon originates from the long-range Coulomb interaction between electrons. It's a kind of quantized charge density wave and exhibits collective electronic oscillations on metal nanostructures, providing an effective mechanism for sub-wavelength light confinement and manipulation.[1] In recent years, plasmon excitation in molecular-level systems has been widely studied, such as plasmon catalysis,[2-4] quantum plasmons,[5,6] and plasmon-moleclue strong coupling.[7] Classical electromagnetic theory cannot reasonably describe the properties of plasmons due to the size limitation, which must be dealt with appropriately by the method of quantum mechanics(QM). The concept of molecular plasmon excitation is analogous to traditional classical research, which recognizes the electronic excitation in molecules similar to plasmon characteristics as molecular plasmon excitation.[8,9] The study of molecular plasmons will extend the unique properties of plasmons to the microscopic field.[10,11] Obvious plasmon excitation modes have been found in sodium[12], gold and silver chains and clusters,[10,11,13,14] olefin chains,[15,16] polycyclic aromatic hydrocarbons(PAHs),[14,17-20] graphene[21] and so forth. The research on the properties of molecular plasmons has affected sensing,[22-26] optoelectronics,[27,28,29,18,19] and optical nanocircuits,[30] and it is expected to increase the optical resolution to the molecular scale.

With the continuous emergence of molecular and cluster devices and systems, the investigation on molecule plasmon excitation and coupling properties become more and more important. Recently there are groups working on molecular/cluster plasmon identifying and properties[13-16,20]. However, when more molecules and clusters come together to become a multi-molecular ensemble, the coupling makes it difficult to judge the plasmon excitation, especially when there is charge transfer or bond connection between molecules may make the ensemble as a new molecule. So the relationship of the plasmonic properties between multiple clusters or molecular interactions and the original individual cluster or molecule is necessary but still unclear. Benefit from inspiration of traditional plasmon strong-coupling[31-33], Tuomas P. Rossi et al. used Cavity Quantum Electrodynamics Density Functional Theory (cQED-DFT) to study the plasmon coupling between aluminum clusters and benzene rings found that as the number of benzene rings increases, the coupling phenomenon becomes obvious[7]. The phenomenon of strong coupling characteristic of plasmons also exists in the system of small molecules and clusters makes us focus on the molecular plasmon hybridization (MPH) of small molecules or clusters. Similar to the hybrid phenomenon of particles-particles in the traditional plasmon research system[34], it's unclear whether there is also hybridization between two molecules with plasmon excitation modes. In response to this question, with the help of traditional molecular plasmon hybridization research methods, the hybridization characteristics are reckoned to be employed to analyze through plasmon excitation mode and charge transition in a small molecular chain attach system. On the basis of accurately determining the molecular plasmon excitation mode (including single-particle excitation and plasmon excitation with electron-electron interaction), the PI [14,20] can be used to identify the central chain molecular system plasmon excitation. For

smaller−sized plasmon systems (single molecules or small clusters), electronic excitation is usually quite complicated and need an effective method to distinguish excitation between single−electron and plasmon. The chain olefin system is not applicable in the existing identification methods [13, 14, 16, 35] Another method of identifying molecular plasmons is utilize electron density descriptors to identify plasmons proposed by Sara Gil−Guerrero et al. in 2019.[36] This method focuses on the electron density related to electronic excitation and does not focus on the excitation energy. They analyze the eigenvalues of the differential density matrix obtained by the TD−DFT/TD−HF theoretical method, and explore the excitation process in the plasmonic excitation. The method of identifying molecular plasmons avoids adoption non−interacting electronic models as a reference system for single−electron excitation in molecular mechanisms with discrete orbital energy distributions, and also avoids different scaling produced multiple calculations in same system as described above. Forthmore, the time−related calculation framework was adopted to correctly determine the systems response of electromagnetic radiation, which is particularly important in small systems. Therefore, TD−DFT calculations are applied to analyze the hybrid properties of the two neutral chain molecular linker systems in detail. TD−DFT has been widely used in molecular plasmons to effectively analyze the electronic excitation in molecules or clusters. TD−DFT usually has two forms: real−time formalism (RT−TDDFT) and linear response formalism (LR−TDDFT). Both the RT−TDDFT and the LR−TDDFT can obtain the Kohn−Sham (KS) electron−hole transition terms to effectively analyze the electronic excitation. Under the real−time propagation framework, the electron−hole contribution is a more detailed quantum mechanism to understand plasmon absorption and can be visualized in the 2D transition contribution maps (TCM). [8] Through the phenomenon of the energy level corresponding to the plasmon mode splits found that the plasmon properties of two connected molecules can be directly explained by the theory of plasmon hybridization. In addition, increase or decrease of charge can regulate molecular plasmon excitation.[17-19] Therefore, we conducted specific research on whether the increase or decrease of charge has an effect on the phenomenon of plasmon hybridization. The research on MPH is helpful to deepen the hybridization properties for molecular dimension and fill the lacunae in the field of molecular plasmons research.

Therefore, the paper focuses on MPH research and the framework as follows: first, identify the plasmonic mode of initial molecules ($C_8H_9NH_2$ and $C_8H_9OH$) use a finite one−dimensional electron gas model; then, analyze the hybridization phenomenon and internal mechanism of three hybrid molecular configuration with plasmon index and TDDFT; finally, showed the controllability of the MPH phenomenon.

The research system is shown in Figure 1. In our calculation system, we selected an olefin chain (carbon number is 8) as the initial structure. Because in the previous research, the carbon atom chain has an obvious plasmon excitation mode[16]. Plasmon hybridization requires two systems with plasmon characteristics to be directly distinguishable. Therefore, we have added different electron−donating groups (−OH and −$NH_2$) at one end of the olefin chain. In addition, we initially selected three coupling configurations, namely head−to−head (HH), head−to−tail (HT) and tail−to−tail (TT).

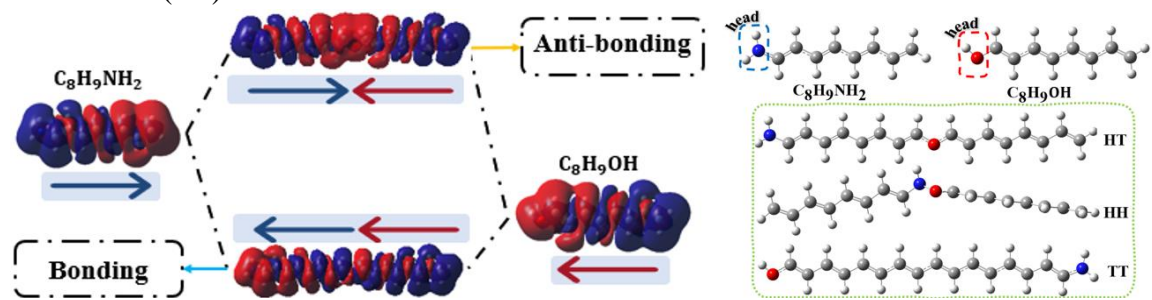

Figure 1 The schematic diagram of plasmon hybridization(left panel) and initial geometry structure of $C_8H_9NH_2$ and $C_8H_9OH$, and the diagram bottom are the hybrid structure of HT, HH and TT configuration, respectively(right panel).

## 2.1 Simulation Methods

The selection of the main skeleton part of the initial structure (i.e., the olefin chain) is based on previous studies [15, 16], and all the theoretical calculations of quantum chemistry are done by Gaussian 16 software package.[37] The DFT calculation[38] for geometric optimization and ground state calculation of the initial structure($C_8H_9NH_2$ and $C_8H_9OH$) use the B3LYP/6−311G(d) method.[39] The ground state geometric optimization and related vibration frequency based on the same level calculation show the minimum value of the optimum potential energy surface and no virtual frequency. The excited state calculation is done by the CAM−B3LYP/6−311G(d) level[40], the D3(BJ) correction[41] are not exploited (no weak interaction involved) for the ground state and excited state calculations. Forthmore, as far as possible consider all the orbital transition contributions in each excited state (greater or equal to $10^{-5}$) to ensure an accurate description of the excited state. In addition, the transition contribution maps (TCM) [42, 43] are used to analyze the plasmon excitation. The calculation of TCM is performed using GPAW software [44, 45] under real−time TD−DFT methods [8, 9, 46]. The calculation used a spatial grid and ensure that there is a 6 angstrom (Å) vacuum area around the molecule. The grid point spacing is 0.3 Å, and the atom is in the center of the space.

In the time−dependent framework (TD−DFT and TD−HF), the relationship between electronic excitation/deexcitation and collective transition can be established through the occupation of virtual electrons. Therefore, the total electron promotion number is from the ground state (0) to the excited state (I) can be expressed as

$N^{0\to I} = \sum_{i=N_{occ}+1}^{N_T} \sum_{a=1}^{N_{occ}} \left( X_{ai}^{I^*} X_{ai}^{I} + Y_{ai}^{I} Y_{ai}^{I^*} \right)$, where $N_T$ represents the total number of orbitals, $N_{OCC}$ represents the number of occupied orbitals, $X_{ai}^{I}$ and $Y_{ai}^{I}$ represents the excitation and deexcitation coefficient, respectively. Due to the normalization of the wave function, the sum of the squares of all possible excitation coefficients associated with

single−electron excitation equals one. Meanwhile, the formula also includes deexcitation terms closely related to collective excitation (between 0−1). The larger contribution of the $Y_{ai}^I$ coefficient, the greater the deviation of $N^{0 \to I}$ from the reference value (1.0) (i.e., deviation from single−electron excitation), so use the logarithmic function to define the plasmonicity index as

$$\delta = 10^{N^{0 \to I} - 1} \tag{1}$$

In addition, according to the transition density matrix for an excitation I obtained from TD−DFT (or TD−HF) calculate further deduced the corresponding transition density

$$\rho^I(\mathbf{r}) = \sum_{a=1}^{N_{occ}} \sum_{i=N_{occ}+1}^{N_T} X_{ai}^I \phi_a^*(\mathbf{r}) \phi_i(\mathbf{r}) + Y_{ai}^I \phi_a(\mathbf{r}) \phi_i^*(\mathbf{r}) \tag{2}$$

## 2.2 The one−dimensional finite electron gas model describes the plasmon dispersion relationship

In order to initially analyze the plasmon excitation modes of the two molecules ($C_8H_9NH_2$ and $C_8H_9OH$), we preliminary matched the $C_8H_9OH$ molecular excitation mode calculated by TD−DFT with the one−dimensional(1D) finite electron gas model as shown in Figure 2. In the description of the QM, Stephan Bernadotte et al. proposed a finite−length electron gas model.[13] All wave vectors are quantified due to boundary conditions, because the electron wave function must disappear at the end of the finite line. Therefore, the single−electron excitation energy $\Delta\varepsilon_{sp}(q)$ of a chain with finite length ($L$) can be determined as $\Delta\varepsilon_{sp}(q) \leq \hbar v_F q + \frac{\hbar^2 q^2}{2m}$, $\Delta\varepsilon_{sp}(q) \geq \hbar v_F q - \frac{\hbar^2 q^2}{2m}$ if $q \leq k_F$ and $\Delta\varepsilon_{sp}(q) \geq \frac{\hbar^2 q^2}{2m}$ if $q \geq k_F$. For possible single−electron excitation, the upper limit of the wave vector leads to additional boundaries $\Delta\varepsilon_{sp}(q) \leq \frac{\hbar^2}{m} k_{max} q - \frac{\hbar^2 q^2}{2m}$ if $q > k_{max} - k_F$, where $v_F = \hbar k_F / m$, $k_F = (1/2)a_0^{-1}$, $a_0 = \frac{4\pi\varepsilon\hbar^2}{me^2}$ and $k_{max} = 9(\pi/L)$, this corresponds to a model with 8 carbon atoms positions in a chain of length $L = 8\pi a_0$.

Plasmon excitation originating from zero mode of the dielectric function($\varepsilon$) for a certain frequency ($\omega_{plas}$), in view of the method of calculating the zero mode of $\varepsilon$ in three−dimensional(3D) electron gas model can express the plasmon frequency as a function of the wave vector, thus, the quantized form of the plasmon dispersion of the one−dimensional finite model is obtained

$$\omega_{\text{plas}}^2(q_n) = \frac{2e^2\rho_0}{m}\kappa_0(lq_n)q_n^2 + v_F^2 q_n^2 \qquad (3)$$

where $q_n=(\pi/L)n$ with n=1,2,3,⋯, $\rho_0 = (1/\pi)a_0^{-1}$ and $l = 2a_0$, $\kappa_0$ is the zeroth modified Bessel function of the second kind.[47] The comparison between the excitation modes calculated by TD−DFT and the dispersion curve of the quantum mechanism can preliminarily distinguish the plasmon excitation and single−electron excitation of the molecular chain, and provide an effective basis to accurately identify molecular plasmon excitation modes. The red curve area indicates the single−particle excitation in the one−dimensional finite uniform electron gas model and the electron excitation in this area is independent of the Coulomb interaction. It can be clearly seen from Figure 2a that plasmon excitation (blue solid dots) can be well distinguished from single−particle excitation (blue circle). In addition, the excitation modes analyzed by TD−DFT show that the two lower energy collective excitation modes (P1−P2) can be well matched with the plasmon excitation in the electron gas model. The upper panel of Figure 2b shows the transition density isosurfaces of the two lower plasmon excitation modes, where red corresponds to the positive phase and blue corresponds to the negative phase. The transition density of the plasmon has an envelope that gradually increases from 1 to 2. This can be explained by the classic description that the plasmon, as a collective electron excitation, has a form of standing wave density, which is observed in the transition density. Other oscillations are caused by finite size. Similarly, $C_8H_9NH_2$ has two low−energy plasmon excitation modes (see Figure 3a), and this plasmon excitation modes has a good agreement with the 1D finite plasmon excitation model. The upper panel of Figure 3b show the isosurface of the transition density of $C_8H_9NH_2$, It can be seen that the first two excitation modes are similar to $C_8H_9OH$, except that the vibration direction is opposite.

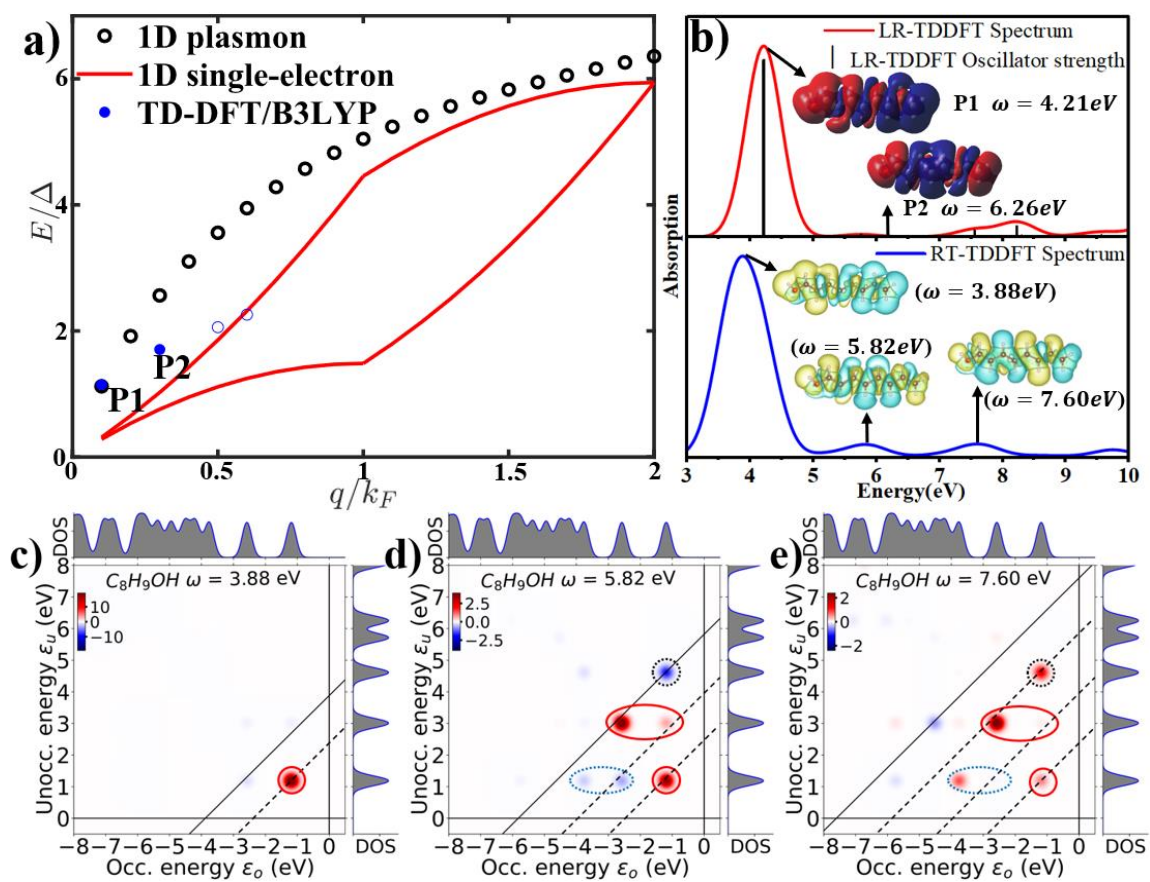

Figure 2 **a**, the plasmon dispersion and particle−hole continuum of an $C_8H_9OH$ chain calculation with TD−DFT compared to the 1D finite−length electron gas model, The excitation vector is shown in unit of the Fermi wave vector $k_F$, where the excitation energy is given relative to the HOMO−LUMO gap. **b**, the LR−TDFDT (upper panel) and RT−TDDFT (lower panel) absorption spectrum of $C_8H_9OH$, the transition density corresponding to two main plasmon excitation modes are visualized in the upper panel and induce density contribution visualized in lower panel. **c−e**, the TCM for the photoabsorption decomposition of $C_8H_9OH$ at different resonance energies ω, the constant transition energy lines $\varepsilon_u-\varepsilon_o=\omega$ are superimposed at the analysis energy (solid line).

In addition, LR−TDDFT and RT−TDDFT were used to analyze the electronic excitations for the both olefin chain molecules. As shown in Figure 2b, the absorption spectra obtained by both theoretical methods have a larger absorption peak located around 4 eV, probably due to the difference in the methods resulting in a smaller red−shift of the optical absorption peak obtained by LR−TDDFT calculations. The first plasmon excitation mode P1 for $C_8H_9OH$ corresponding to the linear response method is located at 4.21 eV, which is in high agreement with the induced density obtained by the real−space method out of 3.88 eV (where yellow corresponds to the positive phase and cyan corresponds to the negative phase), which to some extent indicates the accuracy of the real-space method. Figure 3b also shown the photoabsorption spectrum of $C_8H_9NH_2$, the induced density obtained in real−space located at the resonance energy of 4.35 eV shows a high agreement with the induced density obtained from the linear response, this indicates to a certain extent the accuracy of the time propagation calculation.

The TCM is a useful tool to analyze the response of the Kohn−Sham decomposition which is very useful in reprensenting the plasmon properties, where usually the resonance is a superposition of many electron−hole excitations. Figure 2c shows the TCM as well as the density of states (DOS) of the $C_8H_9OH$ at different resonance energies, where red and represent the positive and negative values of the photoabsorption decomposition in x direction, respectively. Due to the small molecular size, the C8H9OH have well separated discrete KS states, which can also be seen in their DOS diagrams. The TCM corresponding to the C8H9OH molecule have been given and analyzed to find more positive contributions in the lower energy region with increasing resonance energy, accompanied by the generation of negative contributions in response. The TCM of the maximum absorption peak corresponding to the resonance energy ($\omega$ = 3.88 eV) shows a strong positive contribution in the region below the resonance energy, and the low−energy transition that forms this plasmon excitation remains essentially constant within the frequency window as the energy changes. The TCM corresponding to the main resonant frequency of $C_8H_9NH_2$ is shown in Figure 3c, which shows a similarly strong positive contribution at resonant energies of $\omega$=3.70 eV, and its positive contribution increases similarly with increasing energy. Small−molecule plasmon excitation should result from a coupling between the positive contribution of the collective oscillations and the single KS transition. The TCM at the resonance energy of 4.35 eV shows multiple positive contributions, which should be due to s−electron transition. However, for the higher resonance energies (5.70 eV and 7.55 eV) there is a more pronounced negative contribution, which should correspond to single−electron excitations.

In summary, both molecules have two lower energies plasmon excitation modes, and the first excitation mode is strong correspond with a global charge transition.

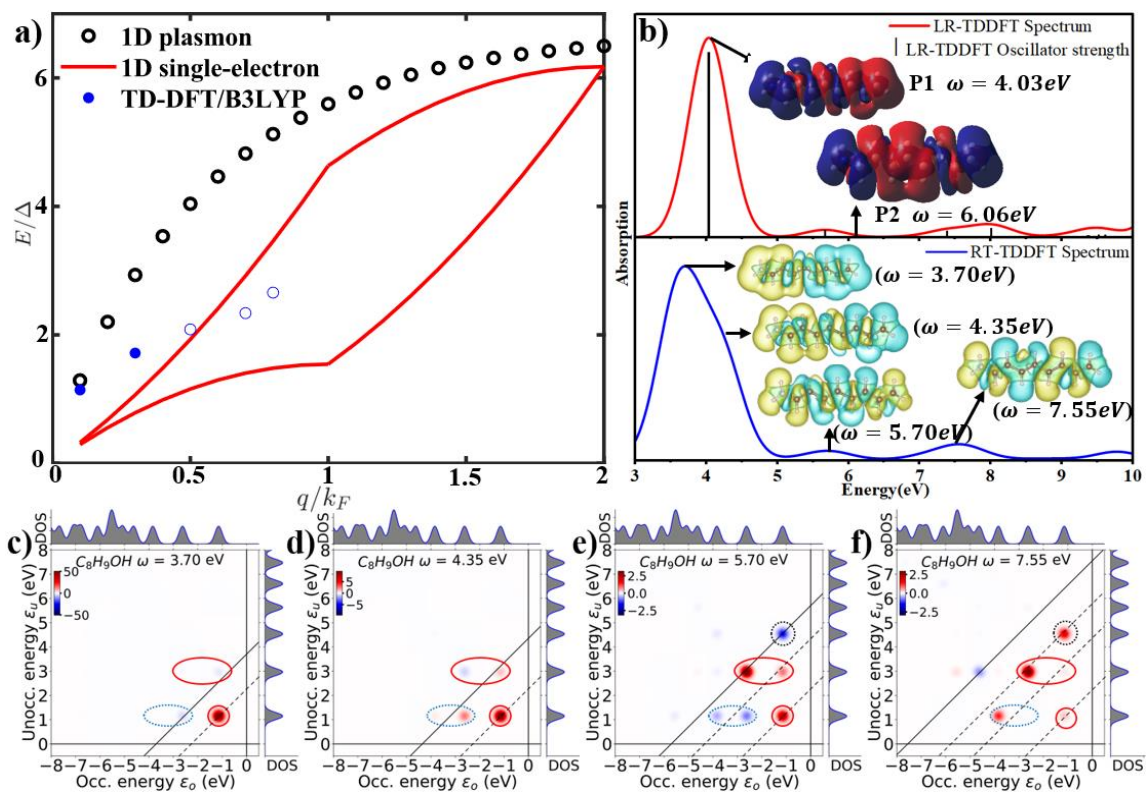

Figure 3 **a**, the plasmon dispersion and particle−hole continuum of an $C_8H_9NH_2$ chain calculation with TD−DFT compared to the 1D finite−length electron gas model, The excitation vector is shown in unit of the Fermi wave vector $k_F$, where the excitation energy is given relative to the HOMO−LUMO gap. **b**, the LR−TDFDT (upper panel) and RT−TDDFT (lower panel) absorption spectrum of $C_8H_9NH_2$, the transition density corresponding to two main plasmon excitation modes are visualized in the upper panel and induce density contribution visualized in lower panel. **c−e**, the TCM for the photoabsorption decomposition of $C_8H_9NH_2$ at different resonance energies ω, the constant transition energy lines $\varepsilon_u-\varepsilon_o=\omega$ are superimposed at the analysis energy (solid line).

## 2.3 Molecular Hybridization Analysis

When the two types of olefin molecules are connected to form a chain, there should be new molecular plasmon modes appear for the excitation. In the following, we will discuss the molecular plasmon modes from hybridization point of view for three different neutral hybrid configurations (HT, HH and TT). The traditional plasmon hybridization research based on hydro electrodynamics theory hold that the plasmon interactions of the hybrid system will cause the plasmon energy shift and the mixing or hybridization between two free plasmon modes (i.e., sphere and cavity plasma)[34]. This hybrid depends on the strength of the interaction and the energy of the original plasmon modes. The coupling and blending of the modes produce the unique tunability of the metal nanoshells. In such plasmon hybridization, usually there is no charge transfer between the two objects but only energy exchanging. However, for the small molecule system, there is charge transfer and in most of conditions new bond forming between the two molecules. Thus the quantum chemical method is different from the classical electromagnetic field. As shown in Figure 4a, the HT configuration shows a strong absorption peak in the low energy region. For $C_8H_9NH_2$ or $C_8H_9OH$, one plasmon excitation mode is found in the strong absorption peak in the low energy region. However, the HT configuration has more than one plasmon modes in the low energy region. Here we only consider the excitation modes with a relatively large PI, and for the excitation modes with a PI almost close to 1, we determine that it is a single−particle excitation through the transition density diagram and ignore it. For the HT configuration, compared with the P1 mode of two single molecules, the mode of the hybrid molecule is split into two modes. The high energy mode P2 (4.28 eV) corresponds to symmetric coupling (anti−bonding) and the low energy mode P1 (3.51 eV) corresponds to antisymmetric coupling (bonding). The charge density distributions in the inset clearly show that it is a hybridization manner for the new plasmon modes. The resonance energy in the coupling system is just like traditional hybridization. For the HH and TT configurations, similar hybrid behavior can be observed (Figure 4b and Figure S1). Compared with the HT configuration, in the hybridization of plasmon excitation modes (P1 and P2) for the HH configuration, the charge transition has changed greatly due to the formation of a certain angle when the two initial molecules are hybridized. For the P1 mode ($S_0 \rightarrow S_1$), the charge transition does not cross the initial molecule but only transitions inside. This may be caused by the offsetting of the opposite charges on the heads of the two initial molecules. For the P2 mode ($S_0 \rightarrow S_2$), the charge transition also occurs inside the initial molecule, and the charge transfer between molecules in integral is weaker. In addition, the charge transfer on the $C_8H_9OH$ molecule is strong and is a collective excitation of the entire molecule. This may be caused by the P1 mode hybridization of the bonding mode of $C_8H_9NH_2$ or $C_8H_9OH$ dipole plasmons, which leads to stronger collective excitation. For P3 ($S_0 \rightarrow S_7$) and P4 ($S_0 \rightarrow S_{10}$) modes, the collective excitation mode occurs on the left and right molecules, respectively, which may be due to coupling is not very strong so that they mainly maintain their own properties. For the TT configuration, it is more complex due to the far distance between the electron−donating groups. For the P1−P5 excitation mode of the TT configuration, it may be formed by the P1 mode hybridization of the initial molecule. The low−energy P1 mode and the high−energy P5 mode both correspond to bonding mode and the P2 and P3 modes correspond to antibonding modes (more details see in Figure S1). There are also many plasmon excitation modes in the high energy region (greater than 4.5 eV), but these plasmon excitation modes are not found in individual $C_8H_9NH_2$ and $C_8H_9OH$. We attribute them to the modes in of the new molecules because in TT configuration, there is just a long carbon chain with no obvious boundary between the two original molecules. However, as the spectra intensity is too weak, they are not discussed any more. The RT−TDDFT results in Figure 4c confirms the hybridization further. The RT−TDDFT utilizing Casida base also shows the electrons and holes correlation very well. The induced electron density in Figure 4c inset also confirms the collective oscillations of the plasmon modes.

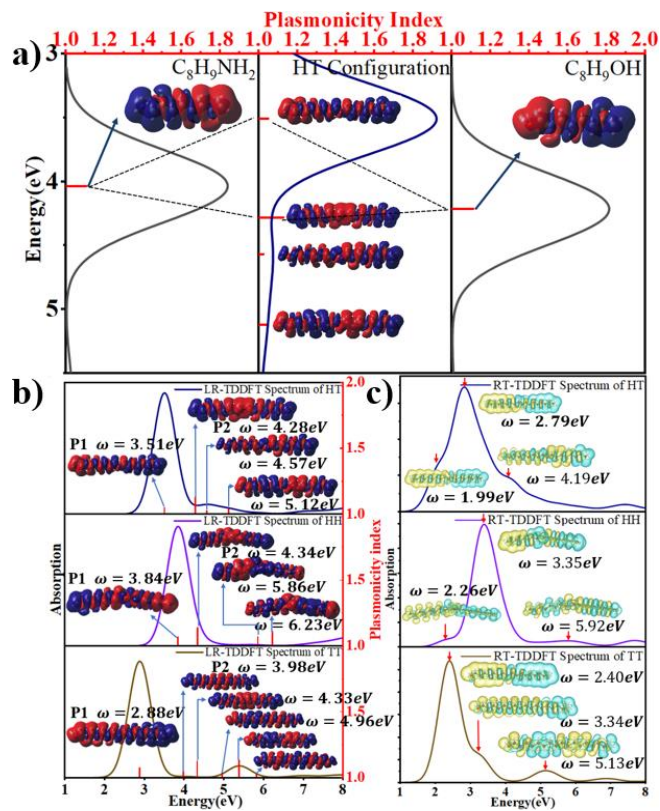

Figure 4 Schematic diagram of plasmon hybridization for HT, HH and TT configuration, respectively. **a.** the hybridization and energy split of HT configuration. **b.** the LR−TDDFT photoabsorption spectrum for HT, HH and TT configuration (curve line), where transition density isosurface of main excitations modes (red vertical line indicates the plasmonicity index) are visualized. **c**. the RT−TDDFT photoabsorption spectrum for HT, HH and TT configuration (curve line), where main resonances are visualized as induce density isosurface.

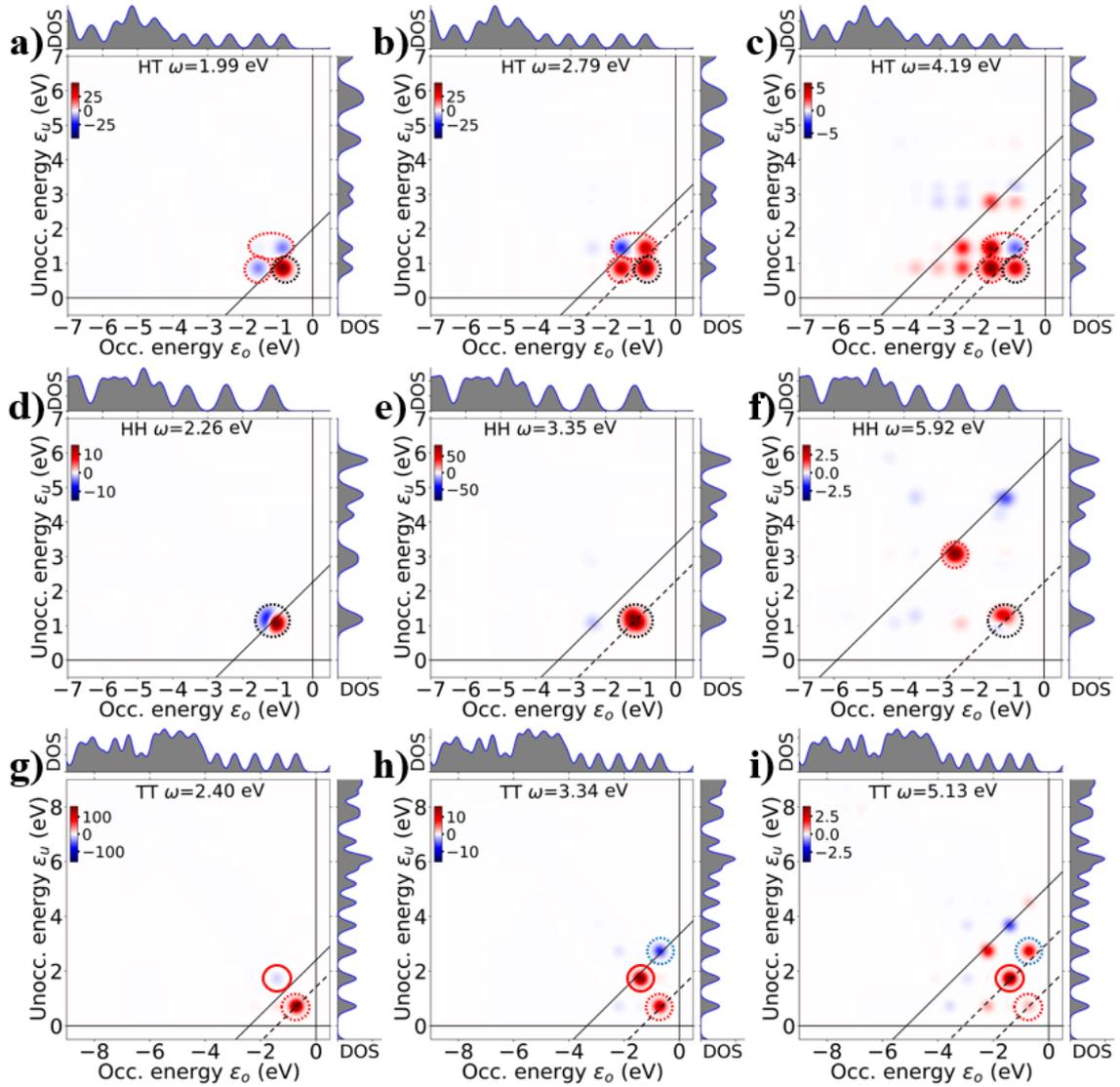

Figure 5  The TCM for the photoabsorption decomposition of HT configuration(a−c), HH configuration(d−f) and TT configuration(g−i) at different resonance energies ω, the constant transition energy lines $\varepsilon_u - \varepsilon_o = \omega$ are superimposed at the analysis energy (solid line).

In addition, the TCM for the photoabsorption spectrum of HT, HH and TT configurations are shown in figure 5. Figure 5a−c shows that the HT configuration gathers more electron−hole contributions of the resonance energy, unlike the more discrete KS contribution of the single molecule, the hybridized HT configuration shows a more concentrated positive contribution for the resonance energy at the highest absorption peak (ω = 2.79 eV), the concentrated and strong positive contribution on the below resonance energy side is important for the formation of plasmon excitations, while the negative contribution on the above resonance energy side may due to the sp electron transition during the plasmon oscillation. For modes with higher resonance energies of plasmon (ω = 4.19 eV), some high−energy electron transition located near the resonance energy (black solid line) are contributing (black elliptical dashed line) additionally to the plasmon excitation. It is worth noting that the positive and negative contributions change below the resonance energy region compared to the positive contribution of ω=2.79 eV (red dashed ellipse), which may be due to the coupling between the positive contribution corresponding to the plasmon and the single−electron transition, which is regarded as the splitting of the plasmon mode into multiple resonances, forming an antisymmetric and symmetric combination with KS transition and plasmon transition. In the existing studies of large−scale architectures, the interaction between plasmons and nearby KS transitions is very weak, and the coupling is regarded as a broadening peak of the plasmon excitation in large cluster structures. Figure 5d−f shows the HH configuration has few and dispersion KS contributions, and as the resonance energy increases the stronger positive contributions separate from the negative contributions and show a larger positive contribution at the resonance energy corresponding to the highest absorption peak (red dashed circles), which may be the reason for the formation of plasmon excitations. Compared to the HT configuration, no significant positive and negative contributions change with frequency, which would lead to weaker hybridization properties in the HH configuration than in the HT configuration. It may be due to the different relative positions of the two different electron−donating groups on the molecular chain: for the HT configuration, the O atom in the middle of the molecule provides a bridge for the overall intramolecular charge transfer and resulting in strong hybridization; while for the HH configuration, both electron−donating groups are located in the middle of the molecular chain. Thus it leads to the formation of local charge transfer on the molecular chain, which may be responsible for the few and dispersion KS contributions. Figure 5d−f shows the TT configuration also appears similarly positive and negative contributions to the HH configuration, but there are relatively more contributions in the region below the resonance energy to produce a stronger positive contribution(red dashed circles) compared to the HH configuration, however, as the energy increases the contribution changes in sign (red realized circles and blue dashed circles), which indicates the coupling in the electron transition which in turn leads to the splitting of the plasmon excitation mode into multiple modes. It is noteworthy that compared to the distribution of the donor electron group in the HT configuration, the two donor electron groups of the TT configuration are located at the ends of the molecular chain that forming just a smooth long chain, which leads to the generation of intramolecular charge transfer similar to those of the HT configuration (the induced electron density at the resonance energy of the highest absorption peak for the HT and TT configurations in Figure 4 show good agreement). In general, the HT configuration is capable of intra−molecular directional strong charge transfer due to the unique arrangement of the electron−donating groups, resulting in a strong concentrated KS electron−hole contribution and a clear split of the plasmon mode into multiple modes, similarly, the TT configuration is capable of intra-molecular charge transfer and leads to hybridization, but for the HH configuration, the intra-molecular charge transfer is relatively weak and accompanied by local electron transfer leading to weaker plasmon hybridization than the other two configuration. The energy level hybridization and splitting diagram of the three configurations are summarized in Figure S2.

### 2.4 Regulatory Molecular Plasmon Hybridization

The plasmons of small molecule systems can be controlled by doping charges.[17] Using this property, we systematically studied the effects of doping negative charges to the hybrid systems on the plasmons excitation. Figure 6 shows the photoabsorption spectrum and induced density of doped charge $C_8H_9OH^-$ and $C_8H_9NH_2^-$ As shown in Figure 6a, the linear response method analysis yields a broad absorption peak in the low energy region of $C_8H_9OH^-$. But there are three excitations with higher PI index and identified them as plasmon excitations. Due to the shorter chain olefin molecule, the value of the PI did not exceed 2. Furthermore, the absorption spectra obtained by the real-space method reflect a good agreement compared to the linear response. For the induced density at the linear response energy of 2.18 eV, it shows good agreement (opposite positive and negative phases) with the induced density at the resonance energy of 2.28 eV in real-space. In addition, we analyzed the induced density at the resonance energy corresponding to the highest absorption peak of the real-space method (3.77 eV), which is consistent with the induced density at the linear response energy of 3.39 eV. The plasmon response of $C_8H_9NH_2^-$ is very similar as shown in Figure 6b.

The TCM for the photoabsorption decomposition of $C_8H_9OH^-$ and $C_8H_9NH_2^-$ at different resonance energies shown in Figure 7. Figure 7b shows the TCM at the resonance energy corresponding to the $C_8H_9OH^-$ maximum absorption peak (3.77 eV) has a large change in KS contribution compared to the neutral molecule (Fig. 2c, ω = 3.88 eV) and shown two strong positive contributions in the region below the resonance energy (black and red dashed circles), in addition, the negative contribution at the resonance energy of 2.27 eV (Figure 7a) changes sign as the energy increases (compare the black dashed circles for the three resonance energies), suggesting that doping charges change the charge transition within the C8H9OH molecule and the KS contribution, which in turn leads to the splitting of the plasmon excitonic mode into multiple modes within the single molecule (see Fig. 6a and b plasmonicity index). For the TCM with a $C_8H_9OH^-$ resonance energy of 6.05 eV (Figure 7c), the two positive contributions to the formation of the plasmon excitation have decayed substantially leading to the excitation not being collective. Similar to neutral monomolecules, the KS contributions of doped-charged $C_8H_9NH_2^-$ (see Figure 7d−f) has a good agreement with $C_8H_9OH^-$, which also indicates that the properties of monomolecules with two different electron−donating groups do not change significantly after doping with charges (see photoabsorption spectrum and PI).

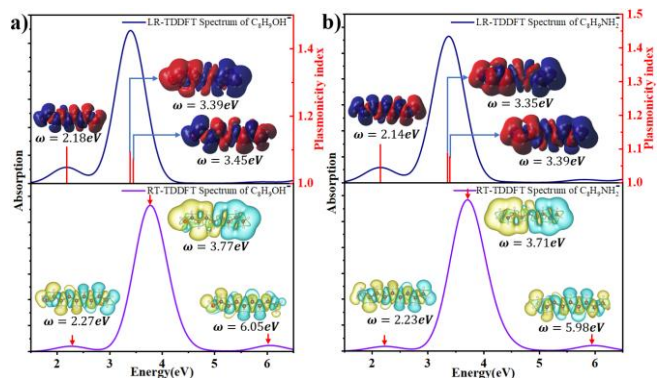

Figure 6  **a.** The LR−TDDFT (upper panel) and RT−TDDFT (lower panel) photoabsorption spectrum of $C_8H_9OH^-$. **b.** The LR−TDDFT (upper panel) and RT−TDDFT (lower panel) photoabsorption spectrum of $C_8H_9NH_2^-$. Where induce density contribution visualized in lower panel.

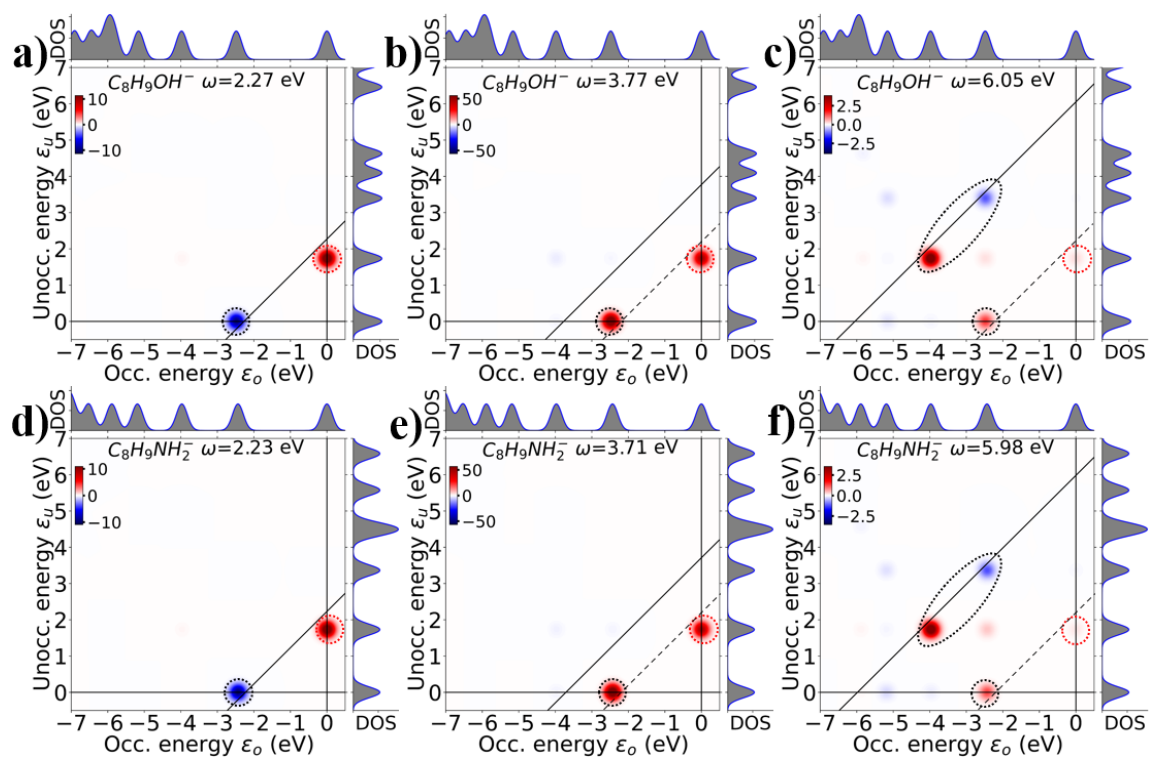

Figure 7  The transition contribution maps for the photoabsorption decomposition of $C_8H_9OH^-$ (a−c) and $C_8H_9NH_2^-$ (d−f) at different resonance energies ω, the constant transition energy lines $\varepsilon_u-\varepsilon_o=\omega$ are superimposed at the analysis energy (solid line).

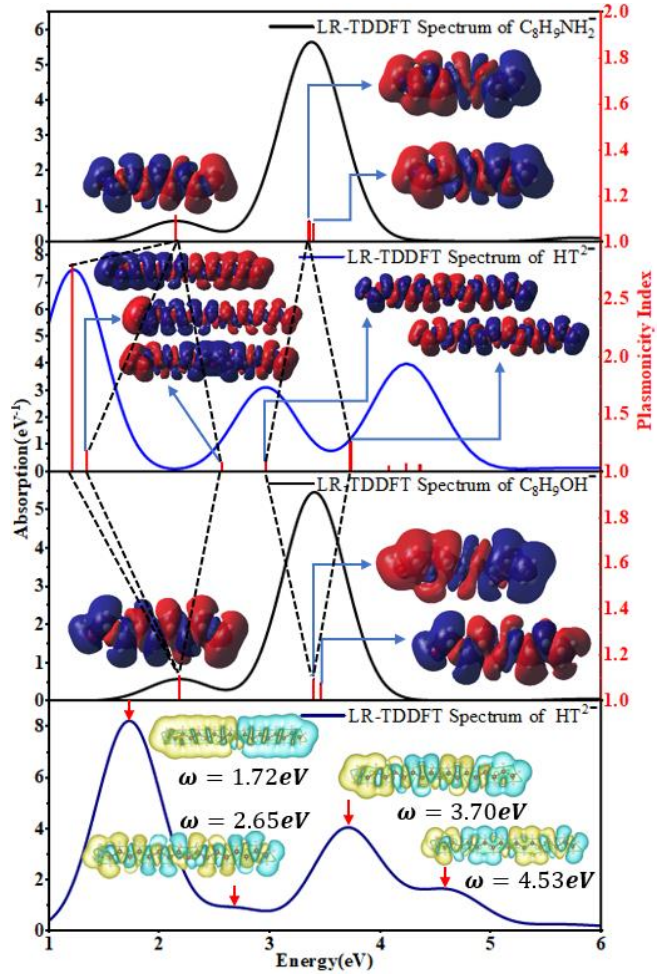

Figure 8 Schematic diagram of plasmon hybridization for $HT^{2-}$ configuration, where transition density isosurface of main excitations modes (red vertical line indicates the PI) are visualized. The lower panel is RT−TDDFT photoabsorption spectrum for $HT^{2-}$ onfiguration (curve line) and the main resonances are visualized as induce density isosurface.

Figure 8 shows the hybridization diagram, plasmonicity index and charge transition isosurface of the HT configuration doped with negative charges. It is found that a small absorption peak appears in the low−energy region after adds a negative charge to the $C_8H_9OH^-$ molecule system, and the PI is also changed due to the negative charge doping (see form Figure 6), and the $C_8H_9NH_2^-$ molecule also has a similar phenomenon. For the $HT^{2-}$ configuration, the doping of two negative charges significantly changes the molecular PI (compared to the HT configuration of Figure 4 without doped charges), and the absorption spectrum has a relatively high value near 1.2 eV due to the doped charge. Large absorption peak, and the corresponding P1 plasmonicity index is also large. By analyzing the charge density isosurface (as shown in Figure 8), we found that the P1 and P2 excitation modes of doping negatively charged HT configuration are due to the hybridization in turn causes the energy level to split between the negatively charged $C_8H_9OH^-$ and $C_8H_9NH_2^-$ molecules. In addition, a similar energy level splitting phenomenon is also found near 3.5eV. From the hybridization levels and surface charges we can see that for such electron doped system, the plasmon hybridization phenomenon is much clearer and simpler compared with the original molecule hybridization. And doping with negative charges can cause the energy level of the low−energy region to be split compared to undoped charges, which is also confirmed in the RT−TDDFT calculations in Figure 8 lower panel. The hybridization of HH, TT configuration is similar in Figure S3 and S4.

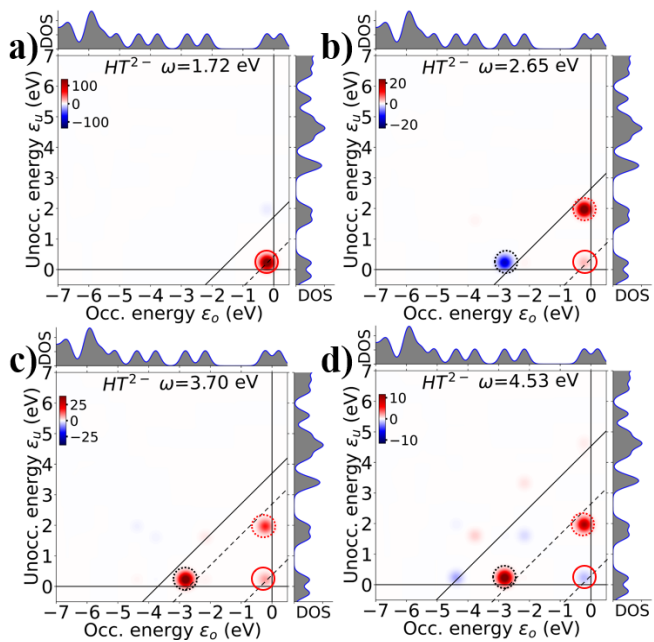

Figure 9 The TCM for the photoabsorption decomposition of $HT^{2-}$ configuration(**a−d**) at different resonance energies ω, the constant transition energy lines $\varepsilon_u-\varepsilon_o=\omega$ are superimposed at the analysis energy (solid line).

In order to deeply analyze the effect of doping charge on electronic excitation as well as hybridization properties, the TCMs doped with two negatively charged $HT_2^-$ configuration and counted in Figure 9. Unlike the neutral HT configuration (see Figure 5a−c), the electron−hole contribution of the doped $HT^{2-}$ configuration is relatively dispersed, and the positive and negative contributions of the TCM at the highest absorption peak resonance energy of 2.56 eV (black and red dashed circles in Figure 9b) show good agreement with the positive and negative contributions of individual $C_8H_9OH^-$ and $C_8H_9NH_2^-$ (see Figure 7a and d black and red dashed circles), which also indicates that the charge transition and electron−hole contribution of the original charged molecule are maintained in the doped $HT^{2-}$ configuration (compare induced density) which is very similar to the traditional hybridization of two different plasmonic objects (i.e. Au and Ag spheres). A similar phenomenon is observed at higher resonance energies (3.70 eV and 4.53 eV) (see Figure 9c and 9d). It is worth mentioning that the KS contributions at all four resonance energies produce an additional positive contribution that gradually decreases with increasing resonance frequency and changes sign to a negative contribution around 4.35 eV, probably due to the formation of a new charge transition after $C_8H_9OH^-$ and $C_8H_9NH_2^-$ hybridization and the accompanying split of a new plasmon excitation mode. The TCM of the $HH^{2-}$ configuration (see [Figure S5a-c](#)) shows a similar phenomenon to the $HT^{2-}$ configuration that the charge−transition of the original molecule is preserved, whereas for the $TT^{2-}$ configuration, the increase resonance energy is accompanied by the generation of more contributions (see [Figure S5d-f](#)) and overall the electronic contributions of the original molecule are not preserved (black dashed ellipse), but the same contributions (red solid circles) as for the first two charged hybrid configurations ($HT^{2-}$ and $HH^{2-}$) are also present. This may be due to the fact that the electron−donating groups of the $TT^{2-}$ configuration are far apart.

Compared the hybridization of the olefin molecules with and without doping charge, one will notice that for the charged molecules it is obvious that the PI of molecule plasmons are much higher than that of the uncharged molecules, and the hybridization phenomenon is much clearer and purer. For the uncharged system, as the electrons are more confined at the atoms, plasmon oscillations are mixed with the single particle excitations. In the HT and HH configurations, as there are clearly differences for the molecules when they are put together, the hybridization is simpler. But for the TT configuration, as there is no obvious boundary for the two original molecules, the behavior of plasmons is more like just in one molecule with long chain. For the charged molecules, as the extra electrons is not confined just in specific atom, it will cruise in the whole molecule, which also couples electrons in different atoms and the coupling will tend to make the excitations anaphase just like in the coupled oscillators. Therefore, the PI of molecules will be larger. When the molecules coupled together, the cruise electrons will bear more for the plasmon oscillation and the whole system will maintain more original properties of individual molecules. For the HT configuration, the same molecular orientation makes the bonding mode appear easier, which shows the strongest peak. For the HH configuration, the head−to−head orientation makes the antibonding mode much stronger. For the TT configuration, as the carbon chain becomes undistinguishable for the two molecules, it oscillates more like a whole long molecule so the bonding mode is much stronger. It also can be concluded that in the future molecular plasmonic devices, the electron doping is very important as it can be used to tune the plasmon mode easily and the molecules still maintain the original properties.

**Conclusions**

In this work, we have studied the phenomenon of plasmon hybridization in small molecular systems. Using the difference between single−particle excitation (corresponding to the pole of the irreducible response function) and plasmon excitation (corresponding to the zero mode of the dielectric function) in the electron gas model, we use the one−dimensional finite electron gas model to initially identified the plasmon excitation modes for $C_8H_9NH_2$ and $C_8H_9OH$. On this basis, we analyzed the three hybrid configurations (HT, HH and TT) formed by the two initial molecules. Using the plasmonicity index, the isosurface of charge transition density and TCM analysis shown that the HT configuration without doping charges has a better hybridization than the other two configurations. This may be due to the fact that the two electron−donating groups (−$NH_2$ and −OH) are close together and form a strong co−bond effect. Finally, we analyzed the regulation of the molecular plasmon and hybridization phenomenon after doping with negative charge, and found that the doped charge will certain extent affect the absorption spectrum of the initial molecule and the plasmonicity index. Due to this factor, the absorption spectrum of the hybrid configuration doped with two negative charges has also changed, especially the absorption spectrum of the HH configuration doped with charges has a most obvious change (a strong absorption peak), which also leads to a larger change of the plasmonicity index and a more obvious hybridization phenomenon of energy level splitting. Through the analysis of doped charges to regulation molecules plasmon and hybridization, we found that plasmon hybridization is prone to occur on molecular groups with strong conjugated electron−donating groups, and doping charges in this type of configuration can effectively change the molecular plasmon excitation and plasmon hybridization. For those systems that do not have strong electron−donating groups and the electron−donating groups are far away so that unlikely to interact with each other, doping charges will not affect plasmon and hybridization. The above results are helpful for in−depth research in the field of MPH and the development of applications related to MPH.

**Authors contributions**
Y.F. conceived the idea and directed the project. N.G. did DFT calculations. G.Z. and Y.H. partially analyzed the hybridization and contributed helpful discussion. N.G and Y.F analyzed the data and wrote the manuscript. All of the authors revised the manuscript.

**Funding**
This research was supported by the National Natural Science Foundation of China (Grant No. 12074054) and the Fundamental Research Funds for the Central Universities (Grant No. DUT21LK06).

**Conflicts of interest**
The authors declare no competing financial interest.

**Availability of data and material**
The data and material that support the findings of this study are available from the corresponding author upon reasonable request.